\documentclass[conference]{IEEEtran}
\IEEEoverridecommandlockouts

\usepackage{cite}
\usepackage{amsmath,amssymb,amsfonts}

\usepackage{graphicx}
\usepackage{textcomp}
\usepackage{xcolor}

\usepackage{graphicx}
\usepackage{titlesec}
\usepackage{balance}
\usepackage{diagbox}
\usepackage{times}
\usepackage[hyphens]{url}
\usepackage{xr,xfrac,xspace,mathtools,amsmath,amsfonts,amssymb,bm}

\usepackage{amsthm}
\usepackage{thmtools, thm-restate}

\usepackage{epstopdf}
\usepackage{booktabs}
\usepackage{epsfig}

\usepackage{hyperref}

\usepackage[font=small]{subfig}

\usepackage[font=bf]{caption}
\usepackage{float}
\usepackage{cleveref}
\crefformat{section}{\S#2#1#3}
\crefformat{subsection}{\S#2#1#3}
\crefname{equation}{}{}

\usepackage[group-separator={,}]{siunitx}
\usepackage{xcolor,colortbl}
\usepackage{lipsum}
\usepackage{capt-of}
\usepackage{algorithm}

\usepackage{algpseudocode}

\usepackage{multicol}
\usepackage{multirow}
\usepackage[normalem]{ulem}

\usepackage{soul}
\usepackage[utf8]{inputenc}
\usepackage{cite}

\usepackage{relsize}
\usepackage{stmaryrd}
\usepackage{wrapfig}
\usepackage{array}
\usepackage{enumitem}
\usepackage{tikz}
\usetikzlibrary{arrows.meta,calc,backgrounds,angles,quotes}

\makeatletter

\newcommand{\system}{\textsc{multi-Bulyan}\xspace}
\newcommand{\systemA}{\textsc{multi-Krum}\xspace}
\newcommand{\systemB}{\textsc{Bulyan}\xspace}
\newcommand{\medgar}{\textsc{Median}\xspace}

\newtheorem{definition}{Definition}
\newtheorem{theorem}{Theorem}
\newtheorem{lemma}{Lemma}

\newcommand{\nospaceitemize}{\vspace{-0.5\topsep}}

\newenvironment{myalign*}{\par\nobreak\noindent\nonumber\align}{\endalign}

\newcommand{\selectoroffset}{\hspace{-0.07cm}}

\newcommand{\includegraph}[1]{\includegraphics[width=0.85\linewidth,trim={6mm 6mm 6mm 6mm},clip]{figures/\detokenize{#1}.png}}

\newcommand{\complexity}[1]{\ensuremath{\mathcal{O}\!\left({#1}\right)}}
\newcommand{\floor}[1]{\left\lfloor{#1}\right\rfloor}

\allowdisplaybreaks

\newcommand{\norm}[1]{~||#1~||}

\DeclareMathOperator*{\argmin}{arg\,min}
\DeclareMathOperator{\bigO}{\mathcal{O}\selectoroffset}

\makeatother

\def\BibTeX{{\rm B\kern-.05em{\sc i\kern-.025em b}\kern-.08em
    T\kern-.1667em\lower.7ex\hbox{E}\kern-.125emX}}
\begin{document}

\title{Fast and Robust Distributed Learning \\ in High Dimension}

\author{
\IEEEauthorblockN{El-Mahdi El-Mhamdi}
\IEEEauthorblockA{EPFL\\elmahdi.elmhamdi@epfl.ch}
\and
\IEEEauthorblockN{Rachid Guerraoui}
\IEEEauthorblockA{EPFL\\rachid.guerraoui@epfl.ch}
\and
\IEEEauthorblockN{Sébastien Rouault}
\IEEEauthorblockA{EPFL\\sebastien.rouault@epfl.ch}}
\maketitle

\begin{abstract}
Could a gradient aggregation rule (GAR) for distributed machine learning be both robust and fast? This paper answers by the affirmative through
\system. Given $n$ workers, $f$ of which are arbitrary malicious (Byzantine) and $m=n-f$ are not, we prove that  \system can ensure a strong form of Byzantine resilience, as well as an ${\frac{m}{n}}$ slowdown, compared to averaging, the fastest (but non Byzantine resilient) rule for distributed machine learning. When $m\approx n$ (almost all workers are correct), \system reaches the speed of averaging.
We also prove that \system's cost in local computation is $O(d)$ (like averaging), an important feature for ML where $d$ commonly reaches $10^9$, while robust alternatives have at least quadratic cost in~$d$.

Our theoretical findings are complemented with an experimental evaluation which, in addition to supporting the linear $O(d)$ complexity argument, conveys the fact that \system's parallelisability further adds to its efficiency.
\end{abstract}


\section{Introduction}

The ongoing data deluge
has been both a blessing and a burden for machine learning system designers. It was a  blessing since machine learning provably performs better with more training data~\cite{shalev2014understanding}, and a burden since the quantity of available data is huge.
The set of parameters in machine learning is now in the order of a gigabyte~\cite{dean2012large}, while training data is several orders of magnitude beyond that~\cite{dean2012large} and rarely available in the same location. In short, distributed machine learning is not an option anymore; it is the only way to deliver results in a reasonable time for the user.

At the core of the recent success in machine learning lies Gradient Descent (GD). GD is an algorithm that consists in the very simple idea of updating a parameter in the opposite direction of the gradient of a cost function. The parameter tunes the algorithm being trained,
and the cost function typically represents how bad the actual parameter performs. Stochastic Gradient Descent (SGD), a lightweight version of GD, is the workhorse of today's machine learning. When all involved machines are reliable, SGD can be easily distributed. The general recipe is that several machines (called {\it workers}) carry the gradient computation, and a {\it server} aggregates the resulting gradients in order to update the parameter. This setting is now widely called the {\it parameter-server} setting~\cite{li2014scaling} and is the dominant standard in distributed machine learning.

In the mainstream declination of this parameter-server setting~\cite{dean2012large, abadi2016tensorflow,li2014scaling,meng2016mllib}, the  {\it gradient aggregation rule} (GAR) consists mostly in averaging the received gradients. When none of the workers misbehaves (no Byzantine workers) and the system is synchronous, it can be easily proven that averaging the gradients is optimal~\cite{bottou1998online}. Indeed, averaging (or problem-specific variants of it~\cite{zhang2015deep}) requires less steps to train the model by optimizing the use of multiple workers. However, as was recently shown~\cite{krum, xie2018generalized, chen2018draco}, averaging is brittle to the most simple form of adversarial behavior, requiring only one single machine to behave in a malicious manner.

Among the proposed alternatives to averaging, Krum~\cite{krum, el2020robust} attracted significant attention and has been used as a benchmark in most of the body of work on Byzantine-resilient gradient descent (e.g.~\cite{yang2019byrdie, tianxiang2019aggregation, bernstein2018signsgd, xie2018generalized, yang2019bridge, chen2018draco, yang2019adversary, rajput2019detox, munoz2019byzantine, baruch2019little}). Unlike averaging, which incorporates the sum of all proposed gradients (including the Byzantine ones), Krum selects the gradient that is the closest to its $\approx n-f$ neighbors\footnote{More specifically, it is $n-f-2$, and the "$-2$" comes from the fact that each correct worker knows that in its neighbors, there are only $n-f-1$ other workers guaranteed to be correct, besides itself. For safety reasons, the worker exclude 1 worker from its $n-f-1$ neighbourhood, so that the distance to the remaining $n-f-2$ that are selected is  guaranteed to be upper bounded by another \emph{correct} worker.}. This ensures that the model update is performed using a gradient from a worker that behaves correctly.

A key feature of Krum is that, unlike other tools from traditional robust statistics~\cite{rousseeuw1985multivariate}, it does not suffer neither from computability nor from complexity issues. In particular, Krum can be computed in a linear time in $d$, the dimension\footnote{This parameter can attain values as high as $10^9$ or $10^{11}$, thus making any solution from the usual security and fraud detection toolbox unpractical. The latter tends to have a quadratic or even super-quadratic cost as they mostly rely on Principal Component Analysis (PCA) as a key ingredient~\cite{wold1987principal}.} of the model being trained. However, Krum suffers from two important limitations. (1) By basing its selection on a distance measurement, Krum suffers from what is known in high-dimensional machine learning as the {\it curse of dimensionality}. Basically, distances become almost meaningless when the dimension of the vectors is too large. In particular, because models are high dimensional, and the landscape where they are being optimized can be highly non convex, a strong adversary can make a Byzantine resilient GAR converge (as it is proven to do), but to a highly non desirable local minima. Specifically, the adversary benefits from a margin of $\sqrt{d}$ to attack distance-based Byzantine resilient GARs, and make them converge to a bad local minima. (2) By selecting only the best behaving gradient (the closest one to its neighbors), Krum ends up using only one gradient for the update, and does not benefit from the variance reduction that comes from leveraging many workers as in averaging.
Potentially, Krum can have a slowdown of up to $\frac{1}{n}$ compared to using all gradients, when none are Byzantine. (This slowdown is computed as the number of steps it takes averaging to converge, divided by the number of steps it takes Krum to converge.)
Given that $f$ workers are Byzantine, it is understandable that any Byzantine-resilient GAR would have some slowdown compared to averaging, as it keeps supposing that not all $n$ workers are reliable. However, $\frac{1}{n}$ is still too far from a desirable $\frac{n-f}{n}$ (using all $n-f$ non Byzantine workers).

We ask the question, whether we could devise a gradient aggregation rule for distributed SGD that is both robust and fast? We answer by the affirmative through \system{}, which we prove circumvents the two aforementioned limitations.

\system builds on previous work. The intuition behind \system (combining \systemA and \systemB) was sketched in~\cite{aggregathor} without however any precise description or analysis. \systemB~\cite{bulyanPaper} was proposed on top of Krum,  and was proved to divide the attackers leeway by the required $\frac{1}{\sqrt{d}}$ that stems from the so called curse of dimensionality and circumvents the first limitation. For the second limitation, an experimental variant of Krum, coined \systemA was proposed. This does not only take the (single) closest gradient to its $\approx n-f$ neighbors, like Krum does. \systemA goes further and selects the (multiple) $\approx n-f$ closest gradients to their $\approx n-f$ closest neighbors.
However, it was not clear whether (a) \systemA inherits the Byzantine resilience merits of Krum while having a better slowdown of $\approx \frac{n-f}{n}$ and whether (b) \systemB can also be used on top of \systemA to produce the same leeway reduction.

In this paper, we first define two notions of Byzantine resilience, {\it weak} and {\it strong}, in order to take into account the specificity of high dimensional non-convex optimization. We also explain some of the non-intuitive aspects of the latter, such as why "mild" adversarial behavior (noise) can sometimes accelerate learning.
We then prove that \systemA is indeed Byzantine resilient in the same sense as Krum. \systemA  guarantees convergence despite $f$ Byzantine workers. Hence,  \systemB can be used on top of \systemA to guarantee not only convergence, but convergence without the high-dimensional vulnerability of non-convex landscapes. (We call the obtained combination \system.)

We prove (a) that \system ensures strong Byzantine resilience  and  (b) that it is ${\approx \frac{n-2f}{n}}$ times as fast as the optimal algorithm (averaging) in the absence of Byzantine workers. In particular, when $f \ll n$, as is the case in many practical situations~\cite{barroso2013datacenter}, this slowdown tends to $1$ and \system is as fast as the optimal (but non Byzantine resilient) averaging.

\textbf{Paper organization.} In section~\ref{intro}, we present the algorithms being studied,
as well as a toolbox of formal definitions: weak, strong, and $(\alpha,f)$--Byzantine resilience. We also present a necessary context on non-convex optimization, as well as its interplay with the high dimensionality of machine learning together with the $\sqrt{d}$ leeway it provides to strong attackers.
We then introduce some background on non-convex optimization and why it changes the requirement for Byzantine resilient gradient descent. We also formalize our definitions of weak and strong Byzantine resilience (the latter was only informally sketched before).
In Section~\ref{weaksection}, we give our proofs of Byzantine resilience and slow down of \systemA and in Section~\ref{strongsection} our proofs of Byzantine resilience and slow down of \system.
In Section~\ref{sec:experiments}, we evaluate \systemA and \system on two practically important metrics, namely \emph{(1)} the aggregation time achieved on actual hardware and \emph{(2)} the maximum top-1 cross-accuracy reached using a given GAR.
Section~\ref{conclusion} discusses how our work bridges the gap between the Byzantine faults threat model of distributed computing and the data-poisoning threat model of machine learning while proposing new directions.

Our experimental claims could be reproduced using this repository: https://github.com/anonconfsubmit/submit-8618.git with the password 8AS5lds\_lzecbmi95ash.

\section{Background}
\label{intro}

\subsection{Stochastic Gradient Descent}

For illustration, but without loss of generality, consider a learning task that consists in making accurate predictions for the labels of each data instance $\xi_i$ using a high dimensional model (for example, a neural network); we denote the $d$ parameters of that model by the vector $\bm{x}$.
Each data instance has a set of \emph{features} (image pixels), and a set of labels (e.g., \{cat, person\}).
We refer to the $j^{th}$ coordinate of a vector $\bm{x}_i$ with $x_{i,j}$.
The model is trained with the popular backpropagation algorithm based on SGD. Specifically, SGD addresses the following optimization problem.
\begin{equation}
\label{eq:opt}
\min_{\bm{x} \in \mathcal{R}^d} Q(\bm{x}) \triangleq \mathbb{E}_{\xi} F(\bm{x};\xi)
\end{equation}
where $\xi$ is a random variable representing a total of $B$ data instances and $F(\bm{x};\xi)$ is the cost function.
The function $Q(\bm{x})$ is smooth but not convex.

SGD computes the gradient ($\bm{G}(\bm{x},\xi) \triangleq \nabla F(\bm{x};\xi)$) and then updates the model parameters ($\bm{x}$) in a direction opposite to that of the gradient (descent).
The vanilla SGD update rule given a sequence of learning rates $\{\gamma_k\}$ at any given step\footnote{A step denotes an update in the model parameters.} is the following:
\begin{equation}
\label{eq:update}
\bm{x}^{(k+1)} = \bm{x}^{(k)} - \gamma_k \cdot \bm{G}(\bm{x}^{(k)},\xi))
\end{equation}

The popularity of SGD stems from its ability to employ noisy approximations of the actual gradient.
In a distributed setup, SGD employs a \emph{mini-batch} of $b < B$ training instances for the gradient computation:
\begin{equation}
\label{eq:grad}
 \bm{G}(\bm{x},\xi) = \sum_{i=1}^{b}\bm{G}(\bm{x},\xi_i)
\end{equation}

The size of the mini-batch ($b$) induces a trade-off between the robustness of a given update (noise in the gradient approximation) and the time required to compute this update.
The mini-batch also affects the amount of parallelism (Equation~\ref{eq:grad}) that modern computing clusters (multi-GPU etc.) largely benefit from.
Scaling the mini-batch size to exploit  additional parallelism requires however a non-trivial selection of the sequence of learning rates~\cite{goyal2017accurate}.
A very important assumption for the convergence properties of SGD is that each gradient is an unbiased estimation of the actual gradient, which is typically ensured through uniform random sampling, i.e., gradients that are on expectation equal to the actual gradient (Figure~\ref{fig:gradients}).

\label{fig:gradients}

\subsection{Algorithms}

\system relies on two algorithmic components: \systemA \cite{krum} and \systemB \cite{bulyanPaper}.
The former rule requires that $n \ge 2 f + 3$ and the second requires that $n \ge 4 f + 3$. \system (along with \systemA and \systemB used as functions) is presented in Algorithm~\ref{algo}.

Intuitively, the goal of \systemA is to select the gradients that deviate less from the ``majority'' based on their relative distances.
Given gradients $\bm{G}_1 \ldots \bm{G}_n$ proposed by workers 1 to $n$ respectively, \systemA  selects the $m$
gradients with the smallest sum of scores (i.e., $\ell_2$ norm from the other gradients) as follows:

\begin{equation}
\label{winners}
(m)\argmin\limits_{i\in \{1,...,n\}}  \sum\limits_{i \rightarrow j} \|\bm{G}_i - \bm{G}_j \|^2
\end{equation}
where given a function $X(i)$, $(m)\argmin(X(i))$ denotes the indexes $i$ with the $m$ smallest $X(i)$ values, and $i \rightarrow j$ means that $\bm{G}_j$ is among the $n-f-2$ closest gradients to~$\bm{G}_i$.
\systemB in turn takes the aforementioned $m$ vectors, computes their coordinate-wise median and produces a gradient which coordinates are the average of the $m-2f$ closest values to the median.

\setlength{\textfloatsep}{10pt}
\begin{algorithm}[!t]
\caption{The Gradient Aggregation Rule (GAR) of \system.}
\label{algo}
\small
\begin{algorithmic}[1]

\Require $\mathcal{I}$: Set of all items; $\mathcal{U}$: Set of all users.
\Require $n$: number of workers, $m$: minimum \systemA size.

\noindent\underline{\bf Parameter Server:}
\State \noindent{\bf function} \systemA($f, [G_1, ..., G_k]$):
\State \hspace{2mm}$S$ = $Dict()$  \hfill $\rhd$ Dictionary: (key=gradient, value=score).
\State \hspace{2mm}$m = k - f - 2$
\State \hspace{2mm}{\bf for} $i \in [1, ...,k]$ {\bf do}:
\State \hspace{4mm}neighbors = ${k-f-2 \text{ closest (}\ell_2\text{ norm) vectors to }G_i}$
\State \hspace{4mm}$S[G_i] = \sum_{G \in neighbors}\norm{G_i-G}^2$
\State \hspace{2mm}$G^{winner} = S.getSmallestValues(1).getKeys()$
\State \hspace{2mm}$G^{output} = Average(S.getSmallestValues(m).getKeys())$

\State \hspace{2mm}{\bf return} $G^{winner}, G^{output}$
\State \noindent{\bf end function}
\\
\State \noindent{\bf function} \system($f, [G_1, ..., G_n]$):
\State \hspace{2mm}$\theta = n-2f-2$
\State \hspace{2mm}$\beta = \theta-2f$
\State \hspace{2mm}$G^{ext}$ = $Array[\theta][d]$  \hfill $\rhd$  extracted gradients.
\State \hspace{2mm}$G^{agr}$ = $Array[\theta][d]$  \hfill $\rhd$  aggregated gradients.
\State \hspace{2mm}$M$ = $Array[d]$  \hfill $\rhd$  coordinate-wise medians.
\State \hspace{2mm}$C$ = $Array[\beta][d]$
\State \hspace{2mm}{\bf for} $i \in [0, ...,\theta]$ {\bf do}:
\State \hspace{4mm}$G^{ext}[i], G^{agr}[i] =$ \systemA($f, [G_1, ..., G_n] \setminus G^{ext}$)
\State \hspace{2mm}$M$ = $Median(G^{ext})$
\State \hspace{2mm}{\bf for} $j \in [0, ...,d-1]$ {\bf do}:

\State \hspace{4mm}$C[:][j]$ = $Argpartition(|G^{agr}[:][j]-M[j]|, \beta)$ \hfill  $\rhd$  $\beta$ closest coordinates to the median.
\Statex \hspace{2mm}{\bf return $Average(G^{agr}[C])$}
\State \noindent{\bf end function}
\\

\State \noindent{\bf function} GAR($f, [G_1, ..., G_n]$):
\Statex \hspace{2mm}{\bf return} \system($f, [G_1, ..., G_n]$)
\State \noindent{\bf end function}
\end{algorithmic}
\end{algorithm}

\subsection{Byzantine Resilience}

Intuitively, Byzantine resilience requires a $GAR$ to guarantee convergence despite the presence of $f$ Byzantine workers. It can be formally stated as follows.

\begin{definition}[Weak Byzantine resilience]
\label{def:weak}
We say that a $GAR$ ensures weak $f$-Byzantine resilience if the sequence $\bm{x}^{(k)}$ (Equation~\ref{eq:update})
converges almost surely to some $\bm{x^{*}}$ where $\bm{\nabla} Q(\bm{x^{*}})=0$,
despite the presence of $f$ Byzantine workers.
\end{definition}

On the other hand, strong Byzantine resilience requires that this convergence does not lead to ''bad'' optimums, and is related to more intricate problem of non-convex optimization, which, in the presence of Byzantine workers, is highly aggravated by the dimension of the problem as explained in what follows.

\begin{figure}
\centering
\includegraphics[width=0.350\textwidth]{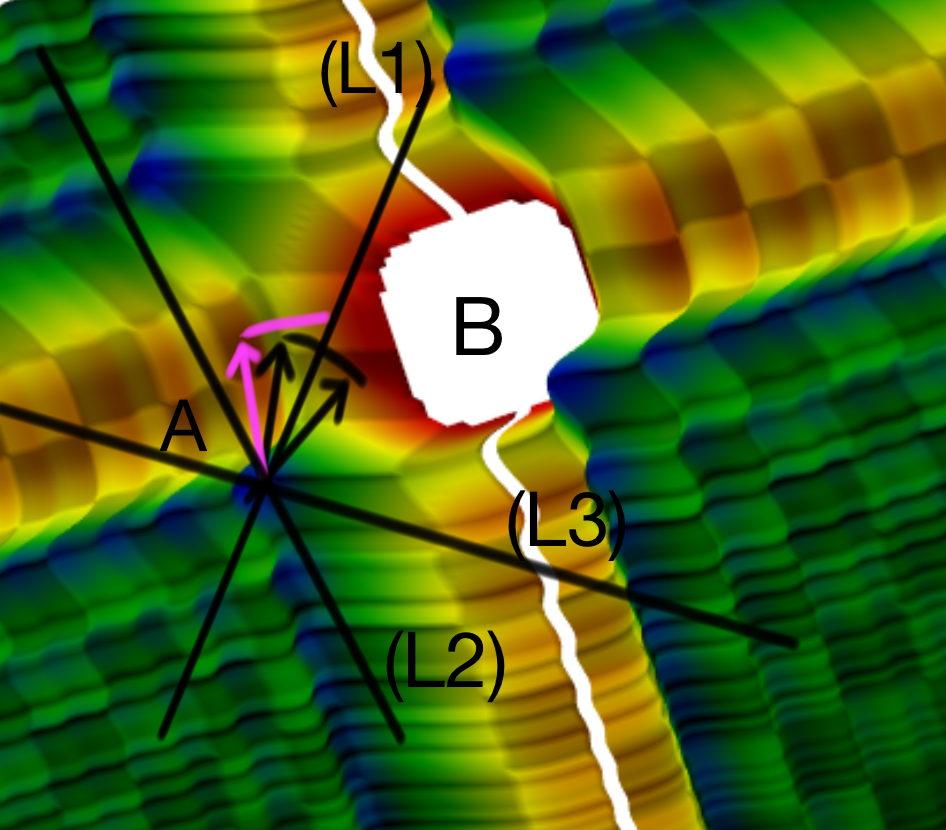}
\caption{In a non-convex situation, two correct vectors (black arrows) are pointing towards the deep optimum located in area B, both vectors belong to the plane formed by lines L1 and L2. A Byzantine worker (magenta) is taking benefit from the third dimension, and the non-convex landscape, to place a vector that is heading towards one of the bad local optimums of area A. This Byzantine vector is located in the plane (L1,L3). Due to the variance of the correct workers on the plane (L1,L2), the Byzantine one has a budget of about $\sqrt{3}$ times the disagreement of the correct workers, to put as a deviation towards A, on the line (L3), while still being selected by a weak Byzantine resilient $GAR$, since its projection on the plane (L1,L2) lies exactly on the line (L1), unlike that of the correct workers. In very high dimensions, the situation is amplified by $\sqrt{d}$.}
\label{timouma}
\end{figure}

\paragraph{Specificity of non-convex optimization.} Non-convex optimization is one of the earliest established NP-hard problems~\cite{haykin2009neural}. In fact, many interesting but hard questions in machine learning have their source in the non convexity of the cost function.

In distributed machine learning, the non-convexity of the cost function creates two non-intuitive behaviours that are important
to highlight.

(1) A "mild" Byzantine worker can make the system converge faster. For instance, it has been reported several times in the literature that noise accelerates learning~\cite{haykin2009neural, bottou1998online}. This can be understood from the "S" (stochasticity) of SGD: as (correct) workers cannot have a full picture of the surrounding landscape of the cost function, they can only draw a sample at random and estimate the best direction based on that sample, which can be, and is probably biased compared to the true gradient. Moreover, due to non-convexity, even the true gradient might be leading to the local minima where the parameter server is. By providing a wrong direction (i.e. not the true gradient, or a correct stochastic estimation), a Byzantine worker whose resources cannot face the high-dimensional landscape of the cost function, might end up providing a direction to get out of that local minima.

(2) Combined with high dimensional issues, non-convexity explains the need for strong Byzantine resilience. Unlike the "mild" Byzantine worker, a strong adversary with more resources than the workers and the server, can see a larger picture and provide an attack that requires a stronger requirement. Namely, a requirement that would cut the $\sqrt{d}$ leeway offered to an attacker in each dimension. Figure~\ref{timouma} provides an illustration.

This motivates the following formalization of strong Byzantine resilience.

\begin{definition}[Strong Byzantine resilience]
\label{def:strong}
We say that a $GAR$ ensures strong $f$-Byzantine resilient if it ensures weak Byzantine resilience and if for every $i \in [1,d]$, there exists a correct gradient $\bm{G}$ (i.e., computed by a non-Byzantine worker) s.t. $\mathbb{E}|\bm{GAR}_i - \bm{G}_i| = O(\frac{1}{\sqrt{d}})$.
The expectation is taken over the random samples ($\xi$ in Equation~1)
 and $\bm{v}_i$ denotes the $i^{th}$ coordinate of a vector~$\bm{v}$.
\end{definition}

\paragraph{Weak vs. strong Byzantine resilience.}

To attack non-Byzantine resilient $GAR$s such as averaging, it only takes the computation of an estimate of the gradient, which can be done in $O(n.d)$ operations per round by a Byzantine worker. This attack is reasonably cheap: within the usual cost of the workload of other workers, $O(d)$, and the server, $O(n.d)$.

To attack weakly Byzantine-resilient $GAR$s however, one needs to find the 'most legitimate but harmful vector possible', i.e one that will (1) be selected by a weakly Byzantine-resilient $GAR$, and (2) be misleading convergence (red arrow in Figure~1).
To find this vector, an attacker has to first collect every correct worker's vector (before they reach the server), and solve an optimization problem (by linear regression) to approximate this harmful but legitimate vector~\cite{bulyanPaper}. If the desired quality of the approximation is $\epsilon$, the Byzantine worker would need at least $\Omega(\frac{n.d}{\epsilon})$ operation to reach it with regression. This is a tight lower bound for a regression problem in $d$ dimensions with $n$ vectors~\cite{haykin2009neural}. In practice, if the required precision is of order $10^{-9}$, with $100$ workers and a neural network model of dimension $10^9$, the cost of the attack becomes quickly prohibitive ($\approx 10^{20}$ operations to be done in each step by the attacker).

To summarize, weak Byzantine resilience can be enough as a practical solution against attackers whose resources are comparable to the server's. However, strong Byzantine resilience remains the only provable solution against attackers with significant resources.

For the sake of our theoretical analysis, we recall the definition of $(\alpha,f)$--Byzantine resilience~\cite{krum} (Definition~\ref{alphaeff}). This definition is a sufficient condition (as proved in~\cite{krum} based on~\cite{bottou1998online}) for  weak Byzantine resilience.
Even-though the property of $(\alpha,f)$--Byzantine resilience is a sufficient, but not a necessary condition for (weak) Byzantine resilience, it has been so far used as the defacto standard~\cite{xie2018generalized, krum} to guarantee (weak) Byzantine resilience for SGD. We will therefore follow this standard and require $(\alpha,f)$--Byzantine resilience from any $GAR$ that is plugged into \system, in particular, we will require it from \systemA. The theoretical analysis done in~\cite{bulyanPaper} guarantees that \systemB inherits it.

Intuitively, Definition~\ref{alphaeff} states that the gradient aggregation rule $GAR$ produces an output vector that lives, on average (over random samples used by SGD), in the cone of angle $\alpha$ around the true gradient. We simply call this the ''correct cone''.

\begin{definition} [$(\alpha,f)$--Byzantine resilience (as in~\cite{krum})]
\label{alphaeff}
Let $0 \le \alpha < \pi/2$ be any angular value, and any integer $0 \le f \le n$.
 Let $V_1,\dots,V_n$ be any independent identically distributed random vectors in $\mathbb{R}^d$, $V_i \sim G$, with $\mathbb{E}G = g$.
 Let $B_1,\dots,B_f$ be any random vectors in $\mathbb{R}^d$, possibly dependent on the $V_i$'s.
 An aggregation rule $GAR$ is said to be $(\alpha,f)$-Byzantine resilient if, for any $1 \leq j_1 < \dots < j_f \leq n$,   vector
 \begin{equation*}
  GAR = GAR(V_1,\dots,\underbrace{B_1}_{j_1},\dots,\underbrace{B_f}_{j_f},\dots,V_n)
 \end{equation*}
 satisfies
    \emph{(i)} $\langle \mathbb{E}GAR, g \rangle \ge (1- \sin\alpha) \cdot \lVert g \rVert^2 > 0$~\footnote{Having a scalar product that is lower bounded by this value guarantees that the $GAR$ of \systemA lives in the aformentioned cone.
    } and
    \emph{(ii)} for $r = 2,3,4$,  $\mathbb{E}~|| GAR ~||^r$   is bounded above by
    a linear combination of terms $\mathbb{E}~|| G ~||^{r_1} \dots \mathbb{E} ~|| G ~||^{r_{n-1}}$
    with $r_1 + \dots + r_{n-1} = r$.
    \label{def:f-byz-resilience}
\end{definition}

\paragraph{Choice of $\bm{f}$.}
The properties of the existing Byzantine-resilient SGD algorithms all depend on one important parameter, i.e., the number of \emph{potentially} Byzantine nodes $f$.
It is important to notice that $f$ denotes a contract between the designer of the fault-tolerant solution and the user of the solution (who implements a service on top of the solution and deploys it in a specific setting).
As long as the number of Byzantine workers is less than $f$, the solution is safe.
Fixing an optimal value for $f$ is an orthogonal problem.
For example, if daily failures in a data center are about 1\

The performance (convergence time) of certain existing Byzantine-resilient SGD algorithms in a non-Byzantine environment is independent of the choice of $f$.
These algorithms do not exploit the full potential of the choice of $f$.
Modern large-scale systems are versatile and often undergo important structural changes while providing online services (e.g., addition or maintenance of certain worker nodes).
Intuitively, there should be a fine granularity between the level of pessimism (i.e., value of $f$) and the performance of the SGD algorithm in the setting with no Byzantine failures.

\section{\systemA: Weak Byzantine Resilience and Slowdown}
\label{weaksection}

Let $n$ be any integer greater than $2$, $f$ any integer s.t $f\leq\frac{n-2}{2}$ and $m$ an integer s.t $m \leq n-f-2$. Let $\tilde m = n-f-2$.\\

Before proving the strong Byzantine resilience of \system, we  first prove the $(\alpha,f)$--Byzantine resilience of \systemA (Lemma~1), then prove its almost sure convergence (Lemma~2) based on that, which proves its weak Byzantine resilience (Theorem~1).

In all what follows, expectations are taken over random samples used by correct workers to estimate the gradient, i.e the "S" (stochasticity) that is inherent to SGD. It is worth noting that this analysis in expectation is not an average case analysis from the point of view of Byzantine fault tolerance. For instance, the Byzantine worker is always assumed to follow arbitrarily bad policies and the analysis is a worst-case one.

The Byzantine resilience proof (Lemma 1) relies on the following observation: given $m\leq n-f-2$, and in particular $m=n-f-2$~\footnote{The slowdown question is an incentive to take the highest value of $m$ among those that satisfy Byzantine resilience, in this case $\tilde m$.}, $m$-Krum averages $m$ gradients that are all in the "correct cone", and a cone is a convex set, thus stable by averaging. The resulting vectors therefore also live in that cone. The angle of the cone will depend on a variable $\eta(n.f)$ as in~\cite{krum}, the value of $\eta(n.f)$ itself depends on $m$. This is what enables us to use $m$ vectors as the basis of our \systemA, unlike~\cite{krum} where a restriction is made on $m=1$. In short, the convexity of a cone allows us to prove lemma~1 almost as in~\cite{krum}, the same derivation applies this time to the averaged $m$ vectors instead of the single vector chosen by Krum.

The proof of Lemma~2 (deferred to the appendix) is the same as the one in~\cite{krum} which itself draws on the rather classic analysis of SGD made by L.Bottou~\cite{bottou1998online}. The key concepts are (1) a global confinement of the sequence of parameter vectors and (2) a bound on the statistical moments of the random sequence of estimators built by the $GAR$ of \systemA. As in~\cite{krum,bottou1998online}, reasonable assumptions are made on the cost function $Q$, those assumption are not restrictive and are common in practical machine learning.

\begin{theorem}[Byzantine resilience and slowdown of \systemA]
\label{theo:weak}
Let $m$ be any integer s.t. $m\leq n-f-2$.
\textbf{(i)} \systemA has weak Byzantine resilience against $f$ failures.
\textbf{(ii)} In the absence of Byzantine workers, \systemA has a slowdown (expressed in ratio with averaging) of $
{{\frac{\tilde m}{n}}}$.
\end{theorem}

\begin{proof}

\textbf{Proof of (i).}
To prove \textbf{(i)}, we will require Lemma 1 and Lemma 2, then conclude by construction of \systemA as an $m$-Krum algorithm with $m = n-f-2$.

\begin{lemma}
Let $V_1,\dots,V_n$ be any independent and identically distributed random $d$-dimensional vectors s.t $V_i \sim G$, with $\mathbb{E}G = g$
  and $\mathbb{E}~|| G - g ~||^2 = d\sigma^2$.
  Let $B_1,\dots,B_f$ be any $f$ random vectors, possibly dependent on the $V_i$'s.
  If $2f + 2 < n$  and $\eta(n,f)\sqrt{d}\cdot \sigma < \lVert g \rVert$,
  where
  \begin{equation*}
    \eta(n,f) \underset{def}{=} \sqrt{2~(n-f + \frac{f\cdot m + f^2\cdot(m+1)}{m}~)}
    ,
  \end{equation*}
  then the $GAR$ function of $\systemA$ is $(\alpha,f)$-Byzantine resilient where $0 \le \alpha < \pi/2$ is defined by
  \begin{equation*}
    \sin \alpha = \frac{ \eta(n,f) \cdot \sqrt{d} \cdot \sigma }{\lVert g \rVert}.
  \end{equation*}
  \label{prop:multikrumresilient}
\end{lemma}

\begin{proof}
  Without loss of generality, we assume that the Byzantine vectors $B_1,\dots,B_f$ occupy the last $f$ positions in the list of arguments of $\systemA$,
  i.e., $\systemA = \systemA(V_1,\dots,V_{n-f},B_1,\dots,B_f)$.

  An index is \emph{correct} if it refers to a vector among $V_1,\dots,V_{n-f}$.
  An index is \emph{Byzantine} if it refers to a vector among $B_1,\dots,B_f$.
  For each index (correct or Byzantine) $i$, we denote by $\delta_c(i)$ (resp. $\delta_b(i)$)
  the number of correct (resp. Byzantine) indices $j$ such that $i \rightarrow j$ (the notation we introduced in Section~3 when defining \systemA), i.e the number of workers, among the $m$ neighbors of $i$ that are correct (resp. Byzantine).

  We have
$
    \delta_c(i) + \delta_b(i) = m$,
      $n-2f-2 \le \delta_c(i) \le m$ and
$          \delta_b(i) \le f.
$

  We focus first on the condition \emph{(i)} of $(\alpha,f)$-Byzantine resilience.
  We determine an upper bound on the squared distance $\lVert \mathbb{E}\systemA - g \rVert^2$.
  Note that, for any correct $j$, $\mathbb{E}V_j = g$.

  We denote by $i_*$ the index of the worst scoring among the $m$ vectors chosen by the \systemA function, i.e one that ranks with the $m^{th}$ smallest score in Equation~\ref{winners}.

  Though we follow the same derivation of~\cite{krum}, one should keep in mind that the manipulated object now is not the winner, minimizing Equation~\ref{winners}, but the ''worst ($m^{th}$) possible vector to choose'' and prove that it also lies in the correct cone. This allows us to average it with the $m-1$ vectors with smaller scores, and by convexity of a cone, prove that the resulting \systemA vector is also in that cone.

  \vspace{-10pt}

  \begin{align*}
  \begin{split}
     ~|| \mathbb{E}\systemA - g ~||^2  & \le ~|| \mathbb{E} ( \systemA \\ & - \frac{1}{\delta_c(i_*)} \sum_{i_* \rightarrow \text{ correct } j} V_j ) ~||^2  \\
         \text{(Jensen inequality)}  & \le \mathbb{E} ~|| \systemA \\ & - \frac{1}{\delta_c(i_*)} \sum_{i_* \rightarrow \text{ correct } j} V_j ~||^2 \\ &
          \le \sum_{\text{correct } i} \mathbb{E} ~|| V_i \\ &- \frac{1}{\delta_c(i)} \sum_{i \rightarrow \text{ correct } j} V_j ~||^2 \mathbb{I}(i_* = i) \\ &
         + \sum_{\text{byz } k} \mathbb{E}~|| B_k \\ & - \frac{1}{\delta_c(k)} \sum_{k \rightarrow \text{ correct } j} V_j ~||^2 \mathbb{I}(i_* = k)
          \end{split}
  \end{align*}
  where $\mathbb{I}$ denotes the indicator function\footnote{$\mathbb{I}(P)$ equals $1$ if the predicate $P$ is true, and $0$ otherwise.}.
  We examine the case $i_* = i$ for some correct index $i$.
  \begin{align*}
    ~|| V_i - \frac{1}{\delta_c(i)} \sum_{i \rightarrow \text{ correct } j} V_j ~||^2 &= ~|| \frac{1}{\delta_c(i)} \sum_{i \rightarrow \text{ correct } j} V_i - V_j ~||^2 \\
   \text{(Jensen inequality)}  &  \le \frac{1}{\delta_c(i)} \sum_{i \rightarrow \text{ correct } j} ~|| V_i - V_j ~||^2 \\
    \mathbb{E} ~|| V_i - \frac{1}{\delta_c(i)} \sum_{i \rightarrow \text{ correct } j} V_j ~||^2  &\le \frac{1}{\delta_c(i)} \sum_{i \rightarrow \text{ correct } j} \mathbb{E}~|| V_i - V_j ~||^2 \\
      &\le 2d\sigma^2.
  \end{align*}
  We now examine the case $i_* = k$ for some Byzantine index $k$.
  The fact that $k$ minimizes the score implies that for all correct indices $i$
  \begin{align*}
    \sum_{k \rightarrow \text{ correct } j} ~|| B_k - V_j ~||^2
      + \sum_{k \rightarrow \text{ byz } l} ~|| B_k - B_l ~||^2
      \le \\ \sum_{i \rightarrow \text{ correct } j} ~|| V_i - V_j ~||^2
      + \sum_{i \rightarrow \text{ byz } l} ~|| V_i - B_l ~||^2.
  \end{align*}
  Then, for all correct indices $i$
  \begin{align*}
    ~|| B_k - \frac{1}{\delta_c(k)} \sum_{k \rightarrow \text{ correct } j} V_j ~||^2
      &\le \frac{1}{\delta_c(k)} \sum_{k \rightarrow \text{ correct } j} ~|| B_k - V_j ~||^2 \\
      &\le  \frac{1}{\delta_c(k)} \sum_{i \rightarrow \text{ correct } j} ~|| V_i - V_j ~||^2 \\ & + \frac{1}{\delta_c(k)} \underbrace{\sum_{i \rightarrow \text{ byz } l} ~|| V_i - B_l ~||^2 }_{D^2(i)}.
  \end{align*}
  We focus on the term $D^2(i)$. Each correct process $i$ has $m$ neighbors, and $f+1$ non-neighbors.
  Thus there exists a correct worker $\zeta(i)$ which is farther from $i$ than any of the neighbors of $i$.
  In particular, for each Byzantine index $l$ such that $i \rightarrow l$, $~|| V_i - B_l ~||^2 \le ~|| V_i - V_{\zeta(i)}~||^2$. Whence
  \begin{multline*}
    ~|| B_k - \frac{1}{\delta_c(k)} \sum_{k \rightarrow \text{ correct } j} V_j ~||^2
      \\
      \le \frac{1}{\delta_c(k)} \sum_{i \rightarrow \text{ correct } j} ~|| V_i - V_j ~||^2 \\  + \frac{\delta_b(i)}{\delta_c(k)} ~|| V_i - V_{\zeta(i)}~||^2 \\  \text{taking expectation on both sides:} \\
    \mathbb{E} ~|| B_k - \frac{1}{\delta_c(k)} \sum_{k \rightarrow \text{ correct } j} V_j ~||^2
      \le \\ \frac{1}{\delta_c(k)} \cdot (\delta_c(i) \cdot 2d\sigma^2  + \delta_b(i) (  \sum_{\text{correct } j \ne i} \mathbb{I}(\zeta(i) = j) \mathbb{E} ~|| V_i - V_j ~||^2  )  )\\
       \le ~(\frac{\delta_c(i)}{\delta_c(k)} \cdot + \frac{\delta_b(i)}{\delta_c(k)} (m+1)~) 2d\sigma^2 \\
      \le ~(\frac{m}{n-2f-2} \\   + \frac{f}{n-2f-2} \cdot (m+1)~) 2d\sigma^2.
  \end{multline*}

  Combining everything we obtain
  \begin{multline*}
    ~|| \mathbb{E}\systemA - g ~||^2  \le \\ (n-f) 2d \sigma^2  + f \cdot ~(\frac{m}{n-2f-2} + \frac{f}{n-2f-2} \cdot (m+1)~) 2d \sigma^2 \\
       \le \underbrace{2~(n-f + \frac{f\cdot m + f^2\cdot(m+1)}{n-2f-2}~)}_{\eta^2(n,f)} d\sigma^2.
  \end{multline*}
  By assumption, $\eta(n,f) \sqrt{d} \sigma < \lVert g \rVert$, i.e., $\mathbb{E}\systemA$ belongs to a ball centered at $g$ with radius $\eta(n,f) \cdot \sqrt{d} \cdot \sigma$. This implies
  \begin{multline*}
    \langle \mathbb{E}\systemA, g \rangle \ge \\ ~( \lVert g \rVert - \eta(n,f) \cdot \sqrt{d} \cdot \sigma  ~) \cdot \lVert g \rVert
      = (1 - \sin \alpha) \cdot \lVert g \rVert^2.
  \end{multline*}
  To sum up, condition~\emph{(i)} of the $(\alpha,f)$-Byzantine resilience property holds.
  We now focus on condition~\emph{(ii)}.
  \begin{multline*}
    \mathbb{E} \lVert \systemA \rVert^r = \\  \sum_{\text{correct } i} \mathbb{E} ~|| V_i ~||^r \mathbb{I}(i_* = i)
                                      + \sum_{\text{byz } k} \mathbb{E} ~|| B_k ~||^r \mathbb{I}(i_* = k)\\
            \le (n-f) \mathbb{E} ~|| G ~||^r + \sum_{\text{byz } k} \mathbb{E} ~|| B_k ~||^r \mathbb{I}(i_* = k).
  \end{multline*}
  Denoting by $C$ a generic constant, when $i_* = k$, we have for all correct indices $i$
  \begin{multline*}
    ~|| B_k - \frac{1}{\delta_c(k)} \sum_{k \rightarrow \text{correct } j} V_j ~||
      \le \\ \sqrt{\frac{1}{\delta_c(k)} \sum_{i \rightarrow \text{ correct } j} ~|| V_i - V_j ~||^2 + \frac{\delta_b(i)}{\delta_c(k)} ~|| V_i - V_{\zeta(i)}~||^2} \\
      \le C \cdot ~( \sqrt{\frac{1}{\delta_c(k)}} \cdot \sum_{i \rightarrow \text{correct } j} ~|| V_i - V_j~|| +  \sqrt{\frac{\delta_b(i)}{\delta_c(k)}} \cdot ~|| V_i - V_{\zeta(i)}~|| ~)\\
      \le C \cdot \sum_{\text{correct } j} ~|| V_j ~|| ~~ \text{(triangular inequality)}.
  \end{multline*}
  The second inequality comes from the equivalence of norms in finite dimension.
  Now
  \begin{multline*}
    ~|| B_k ~||  \le  \\ ~|| B_k - \frac{1}{\delta_c(k)} \sum_{k \rightarrow \text{correct } j} V_j ~|| + ~|| \frac{1}{\delta_c(k)} \sum_{k \rightarrow \text{correct } j} V_j ~|| \\
        \le C \cdot \sum_{\text{correct } j} ~|| V_j ~|| \\
    ~|| B_k ~||^r \le C \cdot \sum_{r_1 + \dots + r_{n-f} = r} ~|| V_1 ~||^{r_1} \cdots ~|| V_{n-f} ~||^{r_{n-f}}.
  \end{multline*}
  Since the $V_i$'s are independent, we finally obtain that $\mathbb{E} ~|| \systemA ~||^r$ is bounded above by a linear combination
  of terms of the form $\mathbb{E} ~|| V_1 ~||^{r_1} \cdots \mathbb{E}~|| V_{n-f} ~||^{r_{n-f}} = \mathbb{E} ~|| G ~||^{r_1} \cdots \mathbb{E}~|| G ~||^{r_{n-f}}$ with $r_1 + \dots + r_{n-f} = r$.
  This completes the proof of condition \emph{(ii)}.
\end{proof}

\label{sec:cvanalysis}

\begin{lemma}
\label{convergence}
Assume that
  \emph{(i)} the cost function $Q$ is three times differentiable with continuous derivatives, and is non-negative, $Q(x) \geq 0$;
  \emph{(ii)} the learning rates satisfy
      $\sum_t \gamma_t = \infty$ and  $\sum_t \gamma_t^2 < \infty$;
  \emph{(iii)} the gradient estimator satisfies
      $\mathbb{E} G(x,\xi) = \nabla Q(x)$ and
      $\forall r \in \{2,\dots,4\},~ \mathbb{E} \lVert G(x,\xi) \rVert^r \le A_r + B_r \lVert x \rVert^r$
    for some constants $A_r,B_r$;
  \emph{(iv)} there exists a constant $0 \le \alpha < \pi/2$
      such that for all $x$
      \begin{equation*}
        \eta(n,f) \cdot \sqrt{d} \cdot \sigma(x) \le \lVert \nabla Q(x) \rVert \cdot \sin\alpha;
      \end{equation*}
  \emph{(v)} finally, beyond a certain horizon, $\lVert x \rVert^2 \geq D$, there exist $\epsilon > 0$ and $0 \le \beta < \pi/2 - \alpha$ such that
    \begin{align*}
      ~|| \nabla Q(x) ~|| &\ge \epsilon > 0 \\
      \frac{ \langle x, \nabla Q(x) \rangle}{\lVert x \rVert \cdot \lVert \nabla Q(x) \rVert} &\ge \cos \beta.
    \end{align*}
    Then the sequence of gradients $\nabla Q(x_t)$ converges almost surely to zero.
    \label{prop:convergence}
\end{lemma}

The proof of Lemma~\ref{convergence} is exactly as in \cite{krum} and is deferred to the appendix.

We conclude the proof of \textbf{(i)} by recalling the definition of \systemA, as the instance of $m-Krum$ with $m=n-f-2$.

\paragraph{Proof of (ii).}
\textbf{(ii)} is a consequence of the fact that $m$-Krum is the average of $m$ estimators of the gradient (line~8 in Algorithm~\ref{algo}). In the absence of Byzantine workers, all those estimators will not only be from the "correct cone", but from correct workers (Byzantine workers can also be in the correct cone, but in this case there are none). As SGD converges in $O(\frac{1}{{m}})$, where $m$ is the number of used estimators of the gradient, the slowdown result follows.
\end{proof}

\section{\system: Strong Byzantine Resilience and Slowdown}
\label{strongsection}

Let $n$ be any integer greater than $2$, $f$ any integer s.t $f\leq\frac{n-3}{4}$ and $m$ an integer s.t $m \leq n-2f-2$. Let $\tilde m = n-2f-2$.\\

\begin{theorem}[Byzantine resilience and slowdown of \system]
\label{theo:strong}
(i) \system provides strong Byzantine resilience against $f$ failures.
(ii) \system requires $O(d)$ local computation.
(iii) When no worker is Byzantine, \system has a ${\frac{\tilde m}{n}}$ slowdown relative to averaging.
\end{theorem}
\begin{proof}
(i) If the number of iterations over \systemA is $n-2f$, then the leeway, defined by the coordinate-wise distance between the output of \systemB and a correct gradient is upper bounded by $\bigO(\frac{1}{\sqrt{d}})$.
This is due to the fact that \systemB relies on a component-wise median, that, as proven in~\cite{bulyanPaper} guarantees this bound.
The proof is then a direct consequence of Theorem~\ref{theo:weak} and the properties of \systemB~\cite{bulyanPaper}. (ii)
The linear cost in $d$ is the consequence of running through the coordinates only in a single loop in \systemB (lines 23-25 in Algorithm~\ref{algo}) and a computation of euclidean distances in \systemA (line~6 in Algorithm~\ref{algo}). Finally, (iii) is a consequence of averaging $\tilde m$ gradients in \system (returned value in line~24 of Algorithm~\ref{algo}).
\end{proof}

\section{Experiments}
\label{sec:experiments}

We report on the performance \system{} (and it component \systemA{}) over two metrics: \emph{(1)} the aggregation time of our implementations of \systemA{} and \system{}, compared to the implementation of \medgar{} in PyTorch, and \emph{(2)} the maximum top-1 cross-accuracy reached on a commonly used classification task in the ML litterature, compared to mere averaging and \medgar{}.

\subsection{Setup}

We run our experiments on the following hardware: \emph{(CPU)} Intel\textregistered{} Core\textsuperscript{\texttrademark{}} i7-8700K @ 3.70GHz, \emph{(GPU)} Nvidia GeForce GTX 1080 Ti, and \emph{(RAM)} $64\text{ GB}$.

We report on the aggregation time, i.e.\ the time needed by a GAR to aggregate its input gradients and provide the output gradient.
This metric is arguably the empirical counterpart of the asymptotic complexity, respectively \complexity{n^2 d}, \complexity{n^2 d} and \complexity{n d} for \systemA{}, \system{} and \medgar{}.
To study the empirical behaviors of \systemA{} and \system{} compared to \medgar{}, we then vary both $n$ and $d$ over a realistic range of values.
Namely we set $\left( n, d \right) \in \left\lbrace 7, 9, 11, \ldots, 35, 37, 39 \right\rbrace \times \left\lbrace 10^5, 10^6, 10^7 \right\rbrace$ and $f = \floor{\frac{n - 3}{4}}$.

\begin{figure}[t]
    \centering
    \includegraph{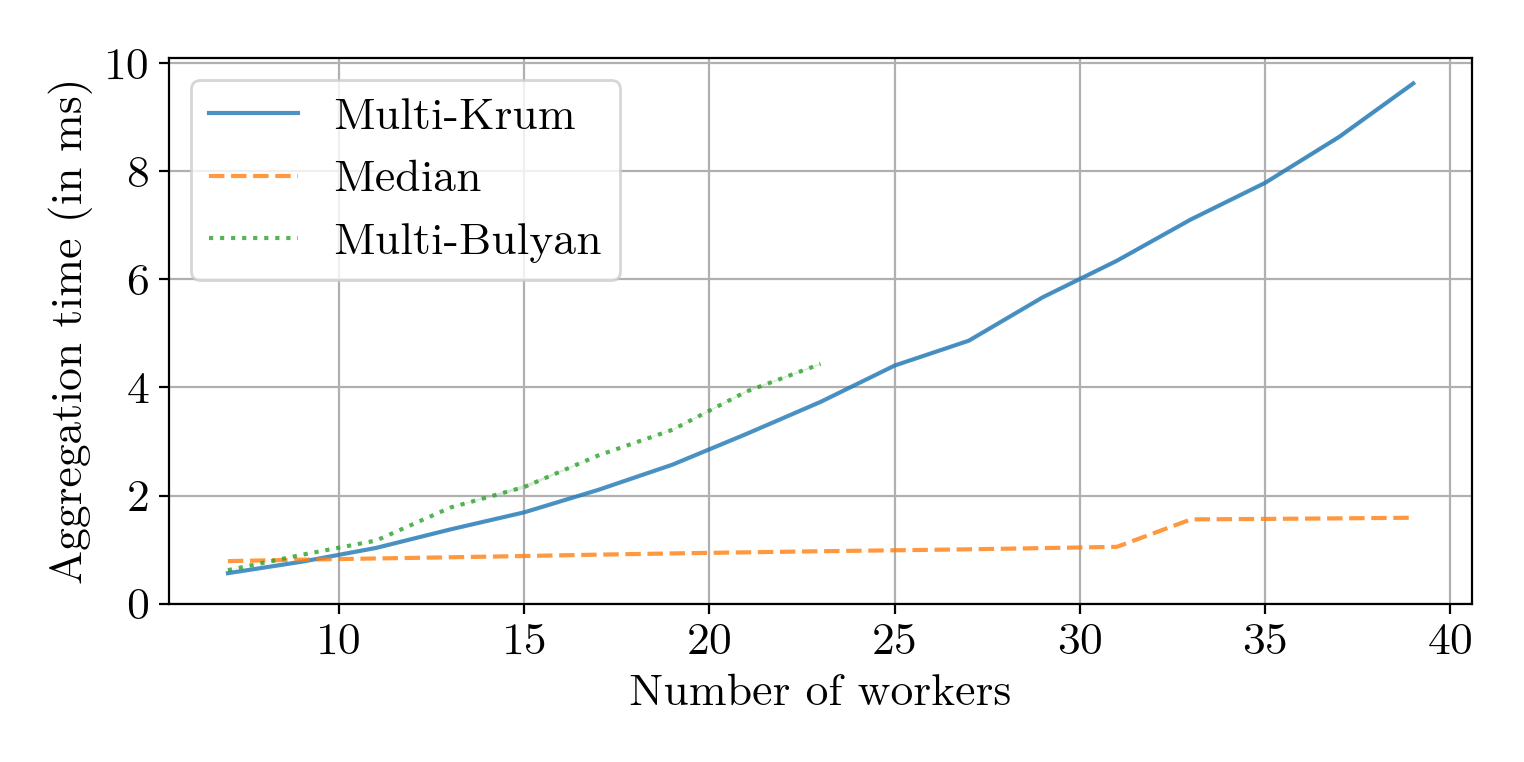}\\[2mm]
    \includegraph{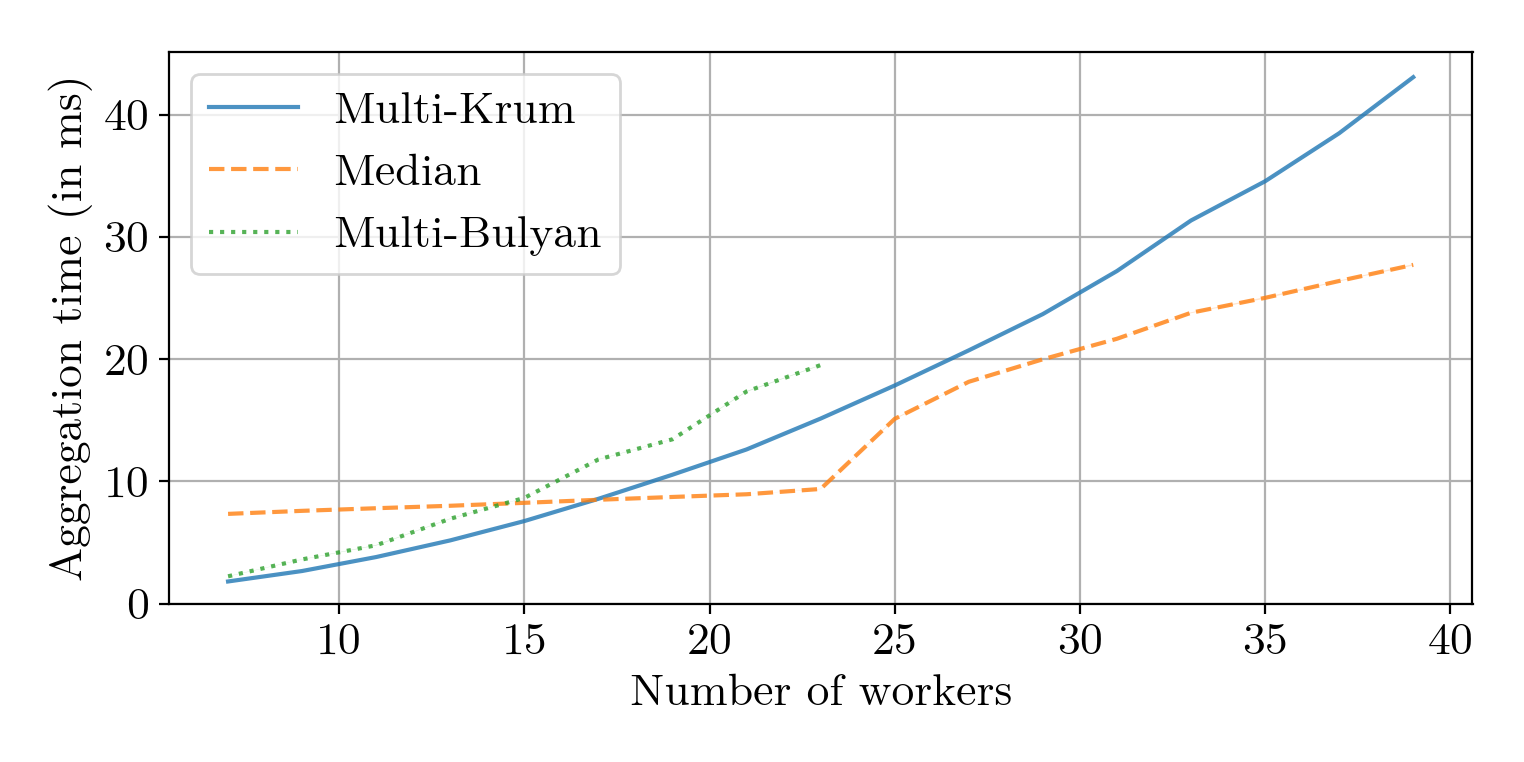}\\[2mm]
    \includegraph{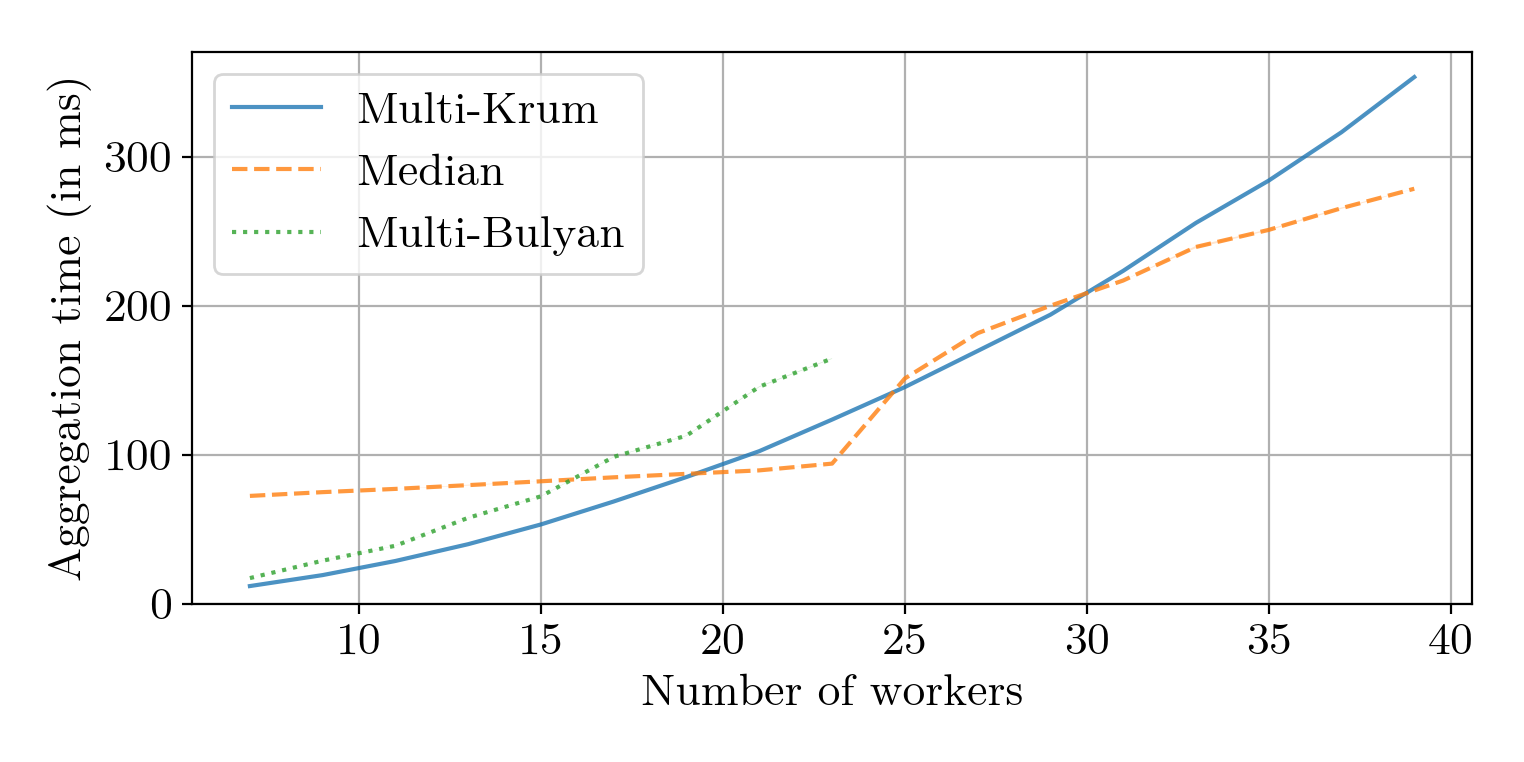}
    \caption{Aggregation time function of the number of aggregated gradients.
    From top to bottom: $\bm{d = 10^5, 10^6, 10^7}$.
    Each experiment is repeated $\bm{7}$ times, and we report on the average and standard deviation of the $\bm{5}$ aggregation times closest to the median.
    The standard deviation in our measurements is very small, barely visible on the graphs.}
    \label{fig:benchmarks}
\end{figure}

The protocol for one run is the following.
$n$ gradients are independently sampled in $\mathcal{U}\!\left( 0, 1 \right)^d$.
These gradients are moved over to the GPU main memory.
The command queue is then flushed on the GPU with \texttt{torch.cuda.synchronize()}, ensuring no kernel is pending on the CUDA stream.
The timer is then started.
The GAR is called on the GPU with the $n$ input gradients.
The command queue is then flushed again, waiting for the GAR's execution to fully complete.
The timer is finally stopped.
There are $7$ runs per values of $\left( n, d \right)$, from which we remove the $2$ furthest execution times from the median of the execution times, and we report on the average and standard deviation of the $5$ remaining measurements in Figure \ref{fig:benchmarks}.

We report on the maximum top-1 cross-accuracy reached by a distributed training process using either \systemA{}, \system{}, \medgar{} or mere averaging for aggregation.
We set $n = 11$ workers and $f = 2$.
There is no attack thought: this experiment highlights the benefits of averaging more gradients per aggregation step, as \systemA{} and \system{} do, over aggregation rules that keep (the equivalent of) only one gradient, e.g.\ \medgar{}.

\begin{figure}[t]
    \centering
    \includegraph{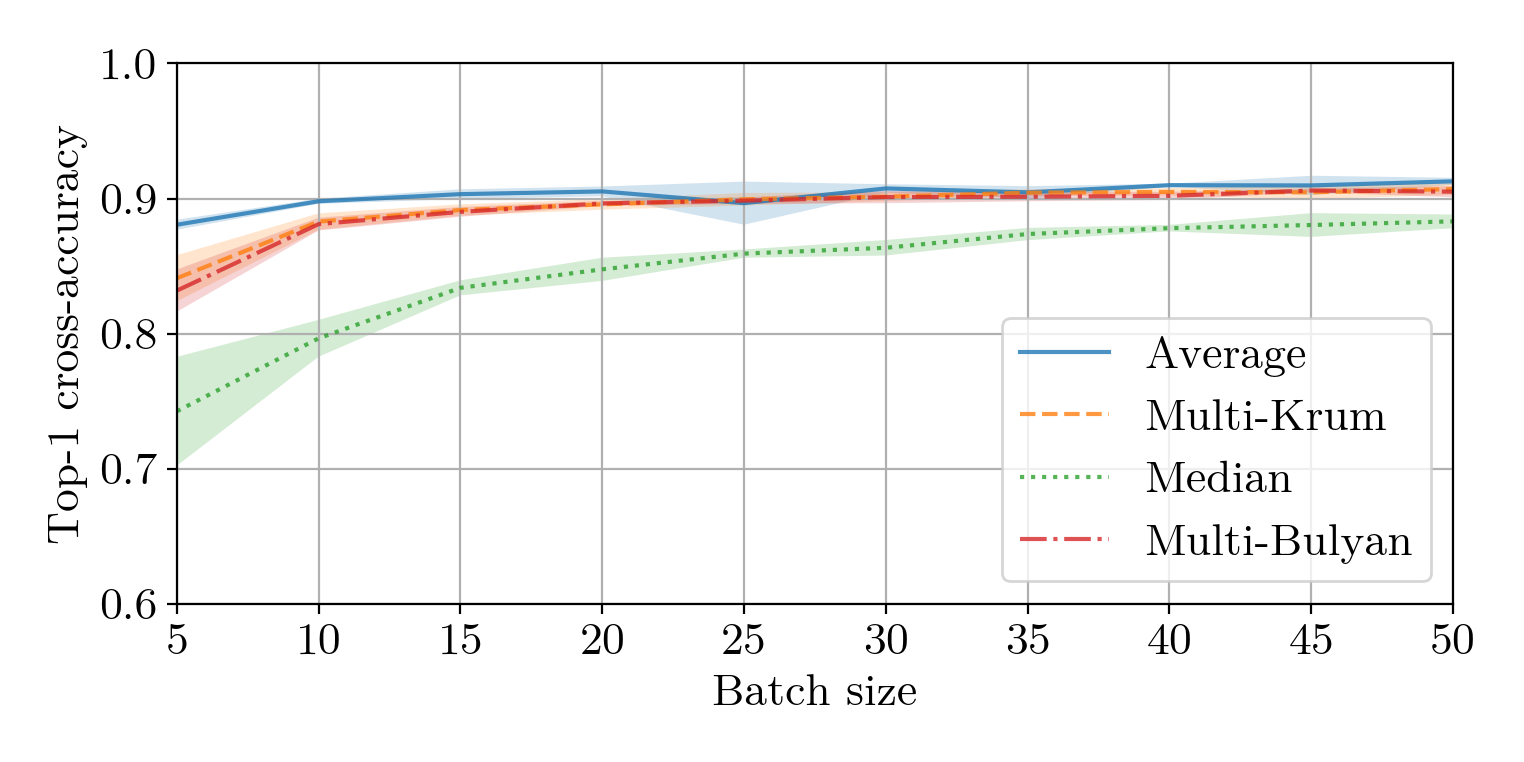}
    \caption{Maximum top-1 cross-accuracy reached by the model with a given GAR and gradient batch size.
    Each experiment is repeated $\bm{5}$ times, with seeds $\bm{1}$ to $\bm{5}$ for reproducibility purpose, and we report on the average and standard deviation of the measured maximum accuracies.}
    \label{fig:max-accuracies}
\end{figure}

The classification task we consider is Fashion-MNIST~\cite{fashionmnist} ($60 000$ training points and $10 000$ testing points).
The model that we train is a convolutional network, composed of two 2D-convolutional layers followed by two fully-connected layers.
The first convolutional layer has $20$ channels (kernel-size $5$, stride $1$, no padding) and the second $50$ channels (same kernel-size, stride and padding).
Each convolutional layer uses the \emph{ReLU} activation function followed by a 2D-maxpool of size $2 \times 2$.
The first fully-connected layer has $500$ hidden units, employing \emph{ReLU}, and the second has $10$ output units.
We train the model using a cross-entropy loss (\emph{log-softmax} normalization + \emph{negative log likelihood} loss) over $3000$ steps, with a fixed learning rate of $0.1$ and momentum $0.9$.
To compute their gradients, each worker employs minibatches of size $b \in \left\lbrace 5, 10, 15, \ldots, 45, 50 \right\rbrace$.
Every $100$ steps we measure the top-1 cross-accuracy of the model over the whole testing set, and we keep the highest accuracy achieved over the whole training.
For reproducibility purpose we seed each training, repeated $5$ times with seeds $1$ to $5$.
We report on the average and standard deviation of the highest accuracy achieved using each GAR and batch size in Figure \ref{fig:max-accuracies}.

\subsection{Experimental Results}

In Figure \ref{fig:benchmarks}, the first observation that we can make is that the computational cost of both \systemA{} and \system{} indeed appears quadratic in $n$, the number of workers.
The number of workers $n$ is kept below $24$ for \system{} due to a limited amount of available on-die \emph{shared memory} on the GPU we used.
Regarding \medgar{}, for which we expect a linear increase with $n$, the tendency is not clear.
The \medgar{} that we used for comparison is provided by the state-of-the-art machine learning framework PyTorch.

In Figure \ref{fig:benchmarks}, and despite a higher asymptotic complexity, \systemA{} and \system{} achieve lower aggregation times than \medgar{} for respectively $n \le 7, n \le 9 \left( d = 10^5 \right)$, $n \le 15, n \le 13 \left( d = 10^6 \right)$ and $n \le 17, n \le 15 \left( d = 10^7 \right)$.
Essentially, the higher the dimension of the model, the higher the number of workers up to which \system{} is more competitive than the \medgar{}.

For reference, ResNet-50 contains $d \approx 24M$ parameters.
For such neural network sizes, major DNN frameworks already show scaling issues when employing only $8$ workers~\cite{parameter-box}.
This inherent limitation the practitioner has to apply on the number of workers not to saturate the standard parameter server (even when using high-throughput $56\text{ Gbps}$ \emph{IP-over-InfiniBand} networks~\cite{parameter-box}) would actually make \systemA{} and \system{} faster than \medgar{} in reasonable deployments (where $n$ is tipically smaller than $20$).
The steady performance of \systemA{} is mostly explained by the fact that its most computationally intensive part, the gradients' pairwise distances computation, is also naturally parallelizable on GPU: it consists in many additions and multiplications executed in parallel.
The remaining computations for \systemA{} merely consists in ordering \emph{scalar} values.
The same applies for \system{}: our implementation does the costly pairwise distance computation only once, and since $f \approx \frac{n}{4}$ the median of \system{} is computed over a substantially reduced set of pre-aggregated gradients.

The empirical ``slowdown'' effect of each GAR is captured in Figure \ref{fig:max-accuracies}.
Each of the studied GAR \emph{throw away} gradients that are, in these experiments, all correct.
Compared to mere averaging the $n = 11$ gradients, aggregating less gradients per step has a tangible impact on the model performance: either more training steps, or higher batch sizes per worker, is needed to compensate.
By averaging only (the equivalent of) one gradient per step, \medgar{} shows in this Byzantine-free settings a tangible loss in top-1 cross-accuracy compared to \system{} and \systemA{}, which both achieve almost the same performance as averaging.
As an additional note, the convolutional model in Figure \ref{fig:max-accuracies} has $d = 431080$ parameters, for $n = 11$ workers (and $f = 2$).
For these settings, both \systemA{} and \system{} also have smaller aggregation times than \medgar{}.

\section{Concluding Remarks}
\label{conclusion}

\textbf{From poisoning to Byzantine faults.} We have proven the Byzantine resilience guarantees of \system and its component \systemA, as well as their slowdown with respect to the fastest (but non Byzantine resilient) gradient aggregation rule: averaging.
We also introduced two notions of Byzantine resilience (weak and strong), which we believe are practically interesting in their own right. The first to guarantee convergence and the second to protect against high-dimensional vulnerabilities.

Our notion of strong Byzantine resilience is robust to all kinds of worker failures:   software bugs,  hardware faults, corrupt data and malicious attacks. In particular, it also encompasses poisoning attacks~\cite{biggio2012poisoning}, an important topic in \emph{adversarial machine learning}.
Mediatised cases of poisoning include social platforms being perturbed by few, acute outlying data-points.
While averaging approaches (and even weakly Byzantine-resilient approaches) are vulnerable to such attacks,
\system{} tolerates them (unless they originate from a majority of users) as reported, e.g.\ in some of the recent work on backdooring federated learning~\cite{xie2020dba}.

We also argue that, even when no obvious network of machines exist, the distributed point of view presented in this paper remains relevant.
What is machine learning if not an attempt to \emph{aggregate} knowledge from distributed sources?
As a concrete instance, an account on a social network posting its own content and interacting (e.g.\ like, comment) with content from other accounts can be seen as a \emph{worker}.
Indeed: each of these account generates data-points that, in turn, produce gradients that can be used to update e.g.\ a recommendation model.
Byzantine-resilient aggregations would then be able to filter out the gradients malicious accounts would be producing.
Through the distributed computing lense, SGD is an algorithm that eventually reaches \emph{agreement} between data sources.

\textbf{The no-free lunch variance requirement.} Since 2017, many alternatives to averaging have been proposed (~\cite{chen2018draco, su2018securing, yin2018byzantine, krum, bulyanPaper, yang2019byrdie, tianxiang2019aggregation, bernstein2018signsgd, yang2019bridge, chen2018draco, yang2019adversary, rajput2019detox, munoz2019byzantine} to list a few). One common aspect underlying all these methods, including ours, is their reliance on some "quality gradient" from the non Byzantine workers. This requirement is expressed in terms of how low should the variance of the gradients be. It is important to note that this requirement is not new in machine learning (it is independent of Byzantine resilience requirements), as an unbounded variance provably prevents convergence~\cite{bottou1998online}.

Recently, attention has been brought~\cite{ baruch2019little} to the hypothesis on bounded variance made in the works that are based on Krum, \systemB, trimmed mean and variants.
This hypothesis was used also in the present work. It is important to note that the limitations that are pointed do not contradict what has been proven in this paper.
Precisely, what we prove is that, as long as the variance of the stochastic gradient is controlled by the norm of the real gradient ($\eta(n,f)\sqrt{d}\cdot \sigma < \lVert g \rVert$), \system keeps making progress (i.e. keeps improving the accuracy of the model).
What has been showed in~\cite{baruch2019little} is that, when the models have converged (and no accuracy gain is made anymore), it is possible to inject erroneous gradients and make them accepted by \system.
This is not a surprise, as the assumption does not hold when convergence has happened (the norm of the gradients becomes close to zero).
In practice, this situation is already prevented by what is called {\it early stopping}~\cite{haykin2009neural}. Because the behavior of SGD can lead to bad models after an excessive number of rounds (due to, among other reasons, the fact that the hypothesis on bounded variance does not hold anymore close to convergence~\cite{bottou1998online}), practitioners tend to have a test-set (different than the training-set) on which the model is tested, when the accuracy on the test-set starts degrading close to convergence, the training is stopped and the previous value of the model is kept.
The attack of~\cite{baruch2019little} would therefore only have been effective if it was preventing SGD from progress in the {\it early steps}, not later and close to convergence.

So far, we have proven that \system is (1) faster than Krum and the Median (the two leading Byzantine resilient GARs), as it relies on \systemA, and (2) more robust than Krum and the Median, as it can make use of \systemB.
An interesting open question is whether any progress can be made on the optimality of the control ratio $\eta(n,f)$ of \systemA and thus of \system? In other words, could we make $\eta(n,f)$ smaller, and still prove the convergence of \systemA and \system, while tolerating even smaller values of the variance? If the answer is negative, are \systemA and \system plugable in the available variance-reducing methods~\cite{haykin2009neural} for SGD?

{\footnotesize \bibliographystyle{acm}
\bibliography{main}}

\newpage

\end{document}